# Transient-Enhanced Surface Diffusion on Native-Oxide-Covered Si(001) Nano-Structures during Vacuum Annealing

H. Lichtenberger, M. Mühlberger, and F. Schäffler[*]

Institut für Halbleiterphysik, Johannes Kepler Universität, A-4040 Linz / Austria

Abstract

We report on the transient-enhanced shape transformation of nano-structured Si(001) surfaces upon *in vacuo* annealing at relatively low temperatures of 900 - 950°C for a few minutes. We find dramatic surface mass transport concomitant with the development of low-energy facets on surfaces that are covered by native oxide. The enhanced surface mass transport ceases after the oxide is completely desorbed, and it is also not observed on surfaces where the native oxide had been removed by HF before annealing.

---

[*] corresponding author: friedrich.schaffler@jku.at

Commercial ultra-large-scale integrated circuits have reached physical gate length well below 100 nm, and self-organization phenomena are explored as an alternative route toward the fabrication of even smaller device structures. However, the fabrication of such small structures is only one precondition for shrinking device dimensions. It is as important to preserve their size, shape and electronic properties during subsequent device processing. Here we concentrate on the shape stability of Si nanostructures during vacuum annealing at around 900°C for a few minutes. Such thermal steps are typically employed for native oxide desorption prior to epitaxial growth, but similar thermal budgets are frequently required during device processing, e.g. after ion implantation. While the shape evolution of structured Si surfaces is well described,[1,2] most of the experimental studies employed long-term, high-temperature anneals. Also, quite often exotic annealing procedures (e.g. a flash to 1200°C [3]) and direct current heating (prone to electro-migration artifacts [4]) were used. Here we employed only cleaning and annealing procedures adapted from standard Si device processes.

The experiments were conducted on 4" floating zone (FZ) Si(001) substrates from Wacker Siltronic with industrial standard specifications. The wafers were diced into 17x17 mm$^2$ samples, and then processed by holographic lithography and reactive ion etching (RIE) in an $SF_6$+$O_2$ plasma. The structures consist of [110] oriented wire arrays of almost rectangular cross section (see Fig. 3 below) with periods of 800 nm, and wire heights of 250 – 300 nm. After RIE the photoresist was stripped in a plasma asher operated at 2.45 GHz. To remove remnants of organic contaminations, the samples were then cleaned in a sequence of organic solvents (trichlorethylen, acetone, methanol), followed by a 5 min treatment in hot $H_2SO_4$:$H_2O_2$ (5:1). Prior to annealing, all samples underwent an RCA cleaning process [5] (SC1 + SC2) with or without a final dip in 4% HF. The RCA process creates a reproducible layer of native oxide with a thickness of < 2 nm. The optional HF dip removes this oxide and provides a hydrogen terminated surface that is stable against oxidation for a few hours.

The annealing experiments were conducted in the ultra high vacuum environment of a Riber SIVA 45 Si molecular beam epitaxy (MBE) apparatus that contains a pyrolytic graphite (PG) element for radiative heating.[6] The samples were mounted contamination-free in all-Si adapter wafers. The annealing cycles correspond to the oxide-desorption routine employed before MBE growth: After sample degassing at 550°C the temperature is ramped up to 900 – 950°C at 230°C/min, kept at this maximum temperature for 1 – 5 min, and is then rapidly quenched to room temperature. The temperature profile is controlled by a calibrated (±15°C) thermocouple located in the radiation zone between the PG heater and the sample.

The samples were characterized before and after annealing on air with a Park Scientific atomic force microscope (AFM) operated in a non-contact mode with Olympus TESP tips. Selected samples were imaged by high-resolution cross-sectional transmission electron microscopy (HRXTEM) at 200 keV with a Jeol 2011 FasTEM.

Fig. 1 shows AFM images of a sample covered with native oxide before (a) and after (b) annealing at 950°C for 4 min. It is obvious that a dramatic morphological change has taken place that converted the originally 300 nm high, almost rectangular wire structures into multiple facetted wires that have lost up to 85% of their original height. The central, triangular part of the wires is made up of {311} facets, whereas the lower part of the cross section is inclined against the [100] direction by about 4.5°.

The experiment was repeated with an identical sample that had the native oxide removed in 4% HF. In that case no noticeable change of the wire cross section was observed after annealing.

To further investigate the relation between oxide desorption and surface diffusion, we studied the shape evolution of the wire arrays. Fig. 2 shows the normalized wire height after annealing versus annealing time for samples with and without a final HF dip. To better resolve the initial stages of annealing, most experiments were conducted at 900°C, but 950°C anneals are also shown for comparison.

Fig. 2 reveals that the wire height of the oxide-covered samples initially decreases with increasing annealing time, and then saturates after about 3 min. The onset of saturation agrees well with the time required for complete oxide desorption at 900°C.[7] The saturation behavior is consistent with the results of the oxide-free samples, which do not show any structural changes. To rule out that possible hydro-carbon contaminations from the HF treatment had interfered with Si surface diffusion, a control experiment was performed: A sample that was annealed after an HF-dip underwent another RCA clean to generated a native oxide layer. Repeating the annealing step on this recycled sample resulted in the same loss of height as expected for an RCA cleaned sample (arrow in Fig. 2). We thus can unambiguously associate the greatly enhanced surface mass transport on Si(001) to the presence of a thin layer of native oxide.

It is also interesting to note that annealing at 950°C leads to a saturated wire height that is significantly lower than the saturation height after a 900°C anneal. Obviously, the activation energies for the oxide desorption mechanism and for the Si mass transport are different.

HRXTEM images with the viewing direction along the [110] oriented wires were recorded to study details of facet formation. Fig. 3(a) shows the cross section of the initial wire structures after RIE: Apart from a slight undercut and some minor rounding with a radius of about 3 nm, the uppermost part of the wires is rectangular, exposing only the (001) and $\{1\bar{1}0\}$ facets. After 60 s at 900°C a top (001) facet still exists, but the former $(1\bar{1}0)$ sidewall facet has been transformed into a lower $(1\bar{1}1)$, and an upper $(3\bar{1}1)$ facet [Fig. 3(b)]. We also found some regions with multiple facet orientations, and transitions regions that cannot be assigned to known low energy facets [Fig. 3(c)].

The general behavior can be very well described by a Monte-Carlo-type simulation[8] that makes only use of the surface energies of stable facets.[1] Under these conditions the $(1\bar{1}0)$ sidewall initially evolves into a $(1\bar{1}1)$ facet which subsequently develops $(3\bar{1}1)$ segments in the upper and lower transition regions to the (001) oriented wire top and trench bottom planes, respectively. Upon further annealing, these two $(3\bar{1}1)$ segments expand on the cost of the $(1\bar{1}1)$ facet, until they finally merge to form a single $(3\bar{1}1)$ facet.

The simulations demonstrate that the shape transformation behaves almost as expected near thermal equilibrium, albeit our annealing temperatures are $\approx 500°C$ below the melting point of Si. Moreover, the greatly enhanced Si diffusion is only observed as long as oxide remains on the surface. Hence, the enhanced surface diffusion and the reaction kinetics of $SiO_2$ desorption are obviously linked. The latter process follows the reaction path:[9]

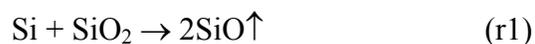

$$Si + SiO_2 \rightarrow 2SiO\uparrow \qquad\qquad (r1)$$

In contrast to $SiO_2$, SiO is volatile at typical annealing temperatures around 900°C. As long as a continuous oxide film exists, SiO forms predominantly at the interface to the oxide.[10] Thermal desorption would then require SiO diffusion through the $SiO_2$ film. It was, however, found that oxide desorption occurs mainly via the expansion of voids that form during an early stage in the oxide.[11,12] Surface diffusion on the Si surface, which becomes exposed within these voids, is high even at 900°C.[13,8] This allows Si transport toward the periphery of the voids, were reaction (r1) and the desorption of SiO can readily occur.

On flat substrates no correlation between void formation and interface structures have been found.[11] Instead, void nucleation has been associated with contaminations on or in the oxide.[14,12] It is, however, not clear whether this applies to our wire-structured substrates, which expose almost atomically sharp intersections between the (001) and {110} facets. Such a large perturbation might be expected to affect the nucleation and anisotropy of void formation. We therefore conducted XTEM decoration experiments to visualize the location of the voids with respect to the wire structure. For that purpose native-oxide covered samples that had undergone an annealing cycle for 1 min at 900°C were covered *in situ* by a Ge layer at a deposition temperature < 175°C. Under these conditions the Ge film becomes amorphous on areas with residual oxide, and polycrystalline in the voids (Fig. 4a). Because of the large mass contrast, the residual native oxide is clearly visible as a lighter stripe. Also, the darker appearance of the poly grains decorates the void regions even in lower resolution images, as can be seen in Fig. 4(b), which shows cross sections of several {311} facetted wires. It is obvious that the voids show no correlation whatsoever with the wire template. It is especially striking that most of the ridges are still completely covered by oxide despite their shape transformation from an originally rectangular cross section. This clearly indicates that the strongly enhanced diffusion occurs predominantly underneath the oxide, and even more, that the oxide follows the shape transformation.

The voids do have some influence, as can be seen at the ridge in the lower left part of Fig. 4(b) that coincides with a void, and has become even flatter than the neighboring oxide-covered ridges. A similar effect might have caused the height fluctuations in Fig. 1(b). Nevertheless, most of the mass transport takes place underneath the oxide, and is most likely associated with the SiO phase that forms with substantial partial pressure[10] at the Si/SiO$_2$ interface upon annealing.

Acknowledgments

We thank C.Schelling for his input during the early stages of this study, and G.Hesser for XTEM preparation. TEM work was performed at the Technical Services Unit (TSE) of the Johannes Kepler University. This work was financially support by GMe and FWF (P12143PHY).

**Figures**

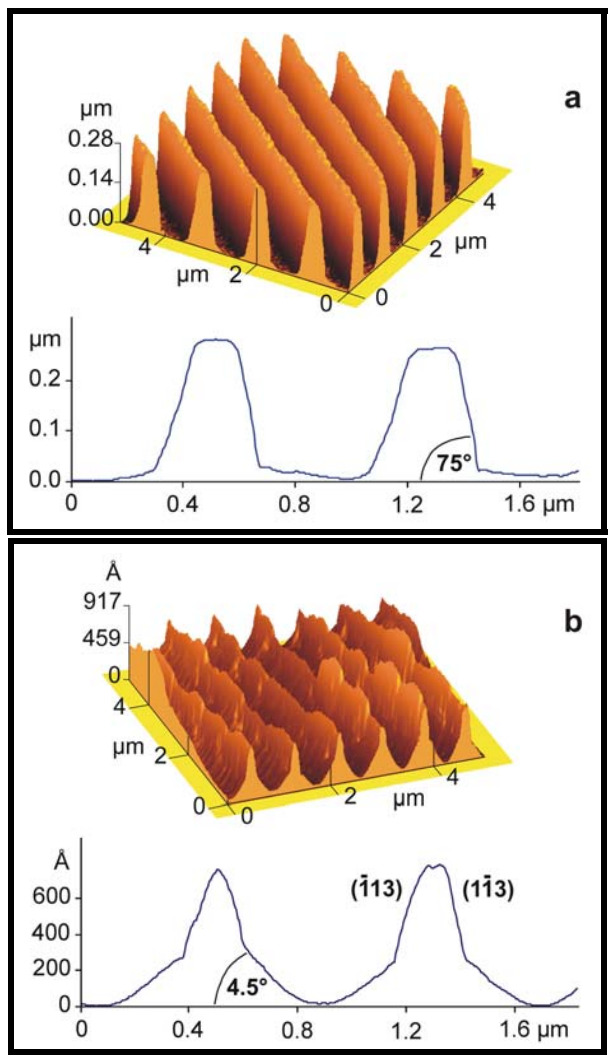

**Figure 1** AFM images and line scans of a native-oxide covered sample before (a) and after (b) annealing at 950°C for 4 min.

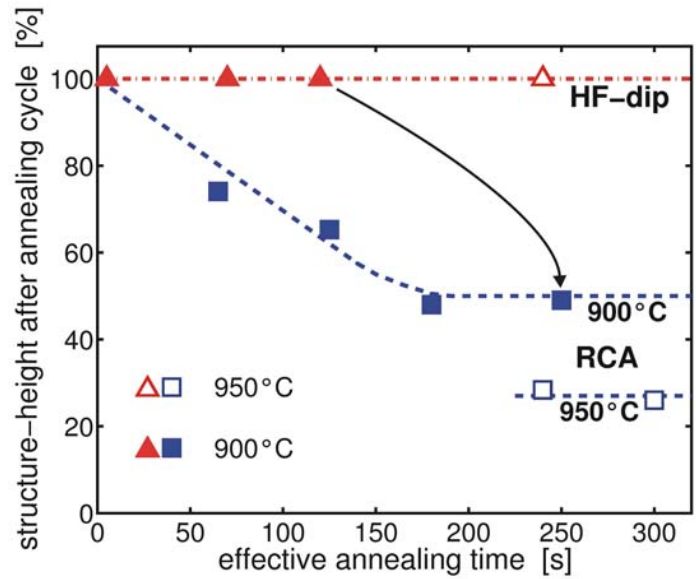

**Figure 2** Normalized wire height vs. annealing time for samples with (RCA), and without (HF) native-oxide. The arrow links the data points from a sample that was annealed first without, and then again with native oxide coverage.

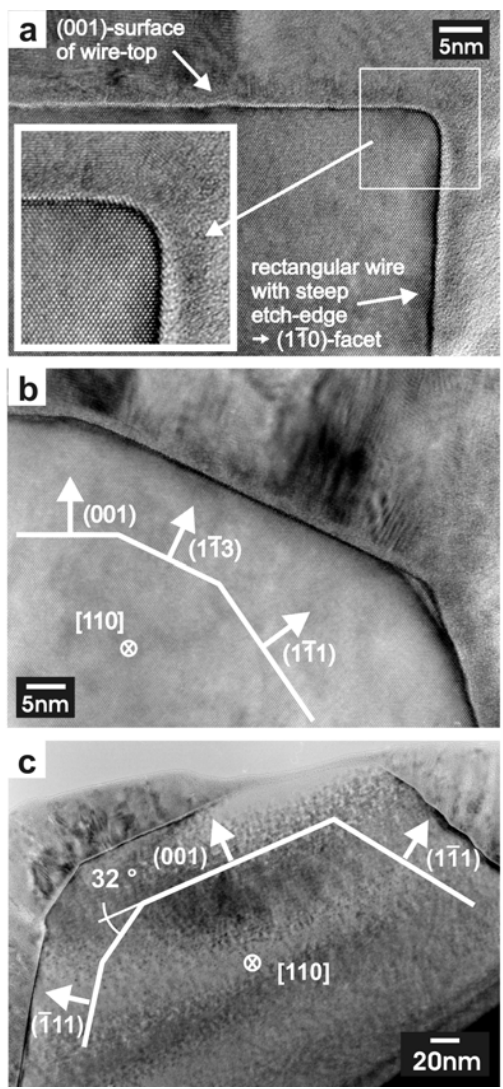

**Figure 3** HRXTEM images of a native-oxide covered sample before (a), and after (b) annealing for 60 s at 900°C; (c) complete wire cross section after annealing.

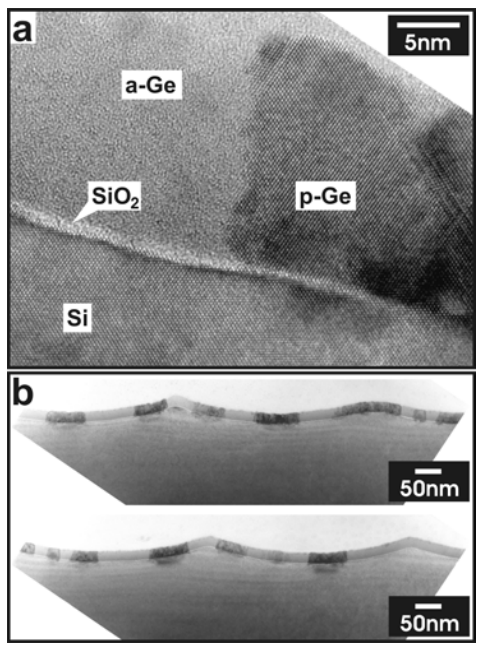

**Figure 4** HRXTEM images of an annealed sample that was *in situ* covered with Ge. (a) Voids in the oxide are decorated by polycrystalline Ge (p-Ge), which appears darker than the amorphous Ge (a-Ge) that forms on $SiO_2$. (b) Lower resolution images of several Si wires.